\documentclass[aps,prd,twocolumn,superscriptaddress,nofootinbib,10pt]{revtex4-1}
\usepackage[colorlinks,linkcolor=blue,anchorcolor=blue,citecolor=blue,urlcolor=blue]{hyperref}
\usepackage{graphicx,subfigure}
\usepackage[figuresright]{rotating}
\usepackage{amsmath,amsfonts,amssymb,bm}
\usepackage{array,enumitem,multirow}
\usepackage{acronym}
\usepackage{cleveref}
\usepackage{makecell}
\usepackage{float}
\newcommand{\be}{\begin{equation}}
\newcommand{\ee}{\end{equation}}
\newcommand{\bea}{\begin{eqnarray}}
\newcommand{\eea}{\end{eqnarray}}
\newcommand{\nn}{\nonumber}
\newcommand{\pd}{\partial}
\newcommand{\td}{\tilde}

\newcommand{\mSun}{{\rm~M_\odot}}
\newcommand{\cO}{{\cal O}}
\newcommand{\TQC}{MOE Key Laboratory of TianQin Mission, TianQin Research Center for Gravitational Physics \&  School of Physics and Astronomy, Frontiers Science Center for TianQin, Gravitational Wave Research Center of CNSA, Sun Yat-sen University (Zhuhai Campus), Zhuhai 519082, China.}

\newacro{GR}{general relativity}
\newacro{GW}{gravitational wave}
\newacro{LVK}{the LIGO-Virgo-Kagra collaboration}
\newacro{CE}{Cosmic Explorer}
\newacro{ET}{Einstein Telescope}
\newacro{LISA}{Laser Interferometer Space Antenna}
\newacro{SNR}{signal-to-noise ratio}
\newacro{FIM}{Fisher Information Matrix}
\newacro{MCMC}{Markov Chain Monte Carlo}
\newacro{RBHC}{relativistic black hole collision}
\newacro{MBH}{massive black hole}
\newacro{CoM}{center-of-mass}
\newacro{CLAP}{close limit approximation approach}
\newacro{PN}{post-Newtonian}
\newacro{ISCO}{Innermost Stable Circular Orbit}
\graphicspath{{./}{./figs/}}
\begin{document}

\title{Detecting relativistic black hole collisions near a massive black hole}

\author{Yirong Fang}
\author{Changfu Shi}
\author{Jianwei Mei}
\email{Email: meijw@sysu.edu.cn}
\affiliation{\TQC}

\date{\today}

\begin{abstract}
Relativistic black hole collisions are one of the most dramatic astrophysical events that can be imagined. 
They could provide the ideal condition for searching for possible new physics beyond general relativity.
However, such events are presumably rare and difficult to occur under normal conditions.
Black holes in a triple system can be accelerated to the relativistic limit and may harbor the chance for a relativistic collision.
In this paper, we study the relativistic black hole collisions in a massive black hole background and the capabilities of several current and future gravitational wave detectors in detecting such signals.
\end{abstract}

\maketitle

\section{Introduction}
\label{sec:intro}

The breakthrough in \ac{GW} detection \cite{Abbott:2016blz} has given us the opportunity to probe the nature of gravity under conditions not possible before \cite{Gair:2012nm,Yunes:2013dva,Berti:2015itd,Yagi:2016jml,Barack:2018yly,Cardoso:2019rvt,Barausse:2020rsu,LISA:2022kgy,Luo:2025ewp,Berti:2025hly}.
However, so far no violation of \ac{GR} has been identified in the detected \ac{GW} signals 
\cite{LIGOScientific:2016lio,LIGOScientific:2019fpa,LIGOScientific:2020tif,LIGOScientific:2021sio,KAGRA:2025oiz,LIGOScientific:2025obp}.
Although the situation may improve with better detection sensitivity in the future, one cannot help wondering what could be the last place to look for possible violations of \ac{GR} should all other searches fail.
In this regard, \ac{RBHC} is the most dramatic astrophysical process imaginable, and they likely have the best condition to reveal possible new physics beyond \ac{GR}.

The collision of black holes has been studied for several decades and has been one of the earliest applications of numerical relativity (see \cite{Price:1994pm} for references).
Past interest in the collision of black holes has come from a variety of topics, such as testing the cosmic censorship conjecture, searching for possible trans-Planckian phenomena at the TeV scale, and so on (see, e.g. \cite{Cardoso:2012qm,Berti:2016rij,Page:2022bem} for reviews and references).
Through numerical simulations, it has been shown that head-on collisions of nonspinning, equal-mass black holes can radiate up to about 13\% of the initial ADM mass in the relativistic limit \cite{Sperhake:2008ga,Healy:2015mla}.
More than half of the radiated energy can be carried by the higher order multipolar components \cite{Ruffini:1973ky,Cardoso:2002ay,Sperhake:2008ga}.
Given the right impact parameter, the radiation efficiency for grazing collisions can be even higher \cite{Pretorius:2007jn,Sperhake:2009jz,Page:2022bem}.
The effect of other parameters, such as spin, charge, and spacetime dimension, has also been studied \cite{Sperhake:2012me,Healy:2022jbh,Healy:2024lhl,Bozzola:2022uqu,Sperhake:2019oaw}.
Further discussions of the nonlinear features involved in the head-on black hole collisions can be found in the recent review \cite{Berti:2025hly} and references therein.

Despite the reoccurring attention and some effort to search for the head-on black hole collision signals in real \ac{GW} data \cite{CalderonBustillo:2020xms,CalderonBustillo:2020fyi,Gayathri:2020coq}, past work has mainly treated \acp{RBHC} as gedanken experiments (see, e.g. \cite{Berti:2010ce,Page:2022bem,Berti:2025hly}).
The obvious reason is that it appears difficult for a free black hole to acquire a large velocity under usual astrophysical conditions.
For example, the fastest stars known in the Galaxy have velocities no greater than 1\% of the speed of light \cite{El-Badry:2023vcd}.
Even under very unusual conditions, the maximum possible astrophysical kick to a black hole is found to be less than 10\% of the speed of light \cite{Healy:2022jbh}.
With such a low initial velocity, the energy efficiency from the head-on collision of black holes is not very high.
It has been found that less than 0.1\% of the total ADM mass will be radiated for the head-on collision of two equal mass non-rotating black holes falling at rest from the spatial infinity  \cite{Davis:1971gg,Anninos:1993zj,Price:1994pm,Anninos:1994gp,Anninos:1995vf}.
This is about two orders below the efficiency of the relativistic collisions \cite{Ruffini:1973ky,Cardoso:2002ay,Sperhake:2008ga,Pretorius:2007jn,Sperhake:2009jz,Healy:2015mla}.

In this paper, we consider the possibility of having \acp{RBHC} in a triple system.
This is inspired by the fact that some of the currently detected \ac{GW} signals may originate from black hole mergers in a \ac{MBH} background (see, e.g. \cite{Bartos:2016dgn,Tagawa:2019osr,Samsing:2020tda,Zhu:2025rdv}).
It is also known that black holes can act as particle accelerators and unbounded \ac{CoM} energy appears to be possible under special conditions \cite{Banados:2009pr,Jacobson:2009zg,Berti:2009bk,Harada:2011xz,Berti:2014lva,Mummery:2025cog}.
It is then natural to consider replacing the colliding particles with black holes, assuming that the \ac{MBH} has a much larger size.
This idea has previously been briefly mentioned, but without a detailed study \cite{Harada:2014vka}.

In the paper, we study how \acp{MBH} can act as accelerators for \acp{RBHC} (for inspiralling \ac{GW} sources around \acp{MBH}, see, e.g. \cite{Chen:2017xbi,Chen:2018axp}).
We find that it is not difficult for black holes orbiting a \ac{MBH} to have relativistic initial boosts.
We also study the capability of several current and future \ac{GW} detectors, including TianQin \cite{TianQin:2015yph,TianQin:2020hid,Luo:2025sos}, LISA \cite{LISA:2017pwj,LISA:2024hlh} (with a similar capability of Taiji \cite{Hu:2017mde}), \ac{CE} \cite{Evans:2021gyd}, \ac{ET} \cite{ET:2019dnz,Branchesi:2023mws} and LIGO \cite{LIGOT2200043}, in detecting the signals of \acp{RBHC} in a \ac{MBH} background.

The paper is organized as follows.
In section \ref{sec:boost}, we study the prospect to get a relativistic speed for black holes moving in a \ac{MBH} background.
In section \ref{sec:wav}, we explain the method used to obtain the \ac{GW} waveform of a symmetric \acp{RBHC}, in which two equal mass and spinless black holes collide with equal and opposite initial boosts.
In section \ref{sec:cap}, we study the capabilities of several current and future detectors to detect these signals.
The paper concludes with a brief summary in section \ref{sec:sum}.

Unless otherwise specified, we use $G_N=c=1$ throughout the paper.

\section{Initial boost}
\label{sec:boost}

The key difference between the collision of black holes and particles in a \ac{MBH} background is that the black holes experience two stages of acceleration:
\begin{itemize}
\item First stage: The acceleration is caused by the \ac{MBH} when the two small black holes reach the collision region.
In this stage, the small black holes can be treated as point particles (The exact mass of the \ac{MBH} is not needed here but we always assume that it is many orders greater than those of the small black holes).
\item Second stage: The acceleration is caused by the mutual gravitational attraction of the two small black holes when they are close enough to each other.
\end{itemize}
In this section, we study the initial boost that the two small black holes can acquire during these two stages of acceleration. 

For the first stage, there has been much study on the motion of particles in a black hole background; see, e.g. \cite{Dymnikova:1986SPU,Poisson:2011nh,Harada:2014vka,Jefremov:2015gza}.
As a proof of principle study, we assume that the two small black holes move in the equatorial plane of the \ac{MBH} with opposite orbital angular momenta, and we further assume that the collision occurs at the periastron of one of the small black holes.
In the following, we recall some basic results on the motion of a test body in the Kerr background, mostly following \cite{Glampedakis:2002ya,Harada:2014vka}.

In the usual Boyer-Lindquist coordinates, the Kerr metric is given by
\bea 
ds^{2}&=&-\Big(1-\frac{2Mr}{\Sigma}\Big)dt^2-\frac{4Mar\sin^2\theta}{\Sigma}dtd\phi\nn\\
&&+\Big(r^2+a^2+\frac{2Ma^2r\sin^2\theta}{\Sigma}\Big)\sin^2\theta d\phi^2\nn\\
&&+\frac{\Sigma}{\Delta}dr^2+\Sigma d\theta^2\,,\label{Oeq:1}\eea
where $\Delta \equiv r^2 - 2Mr +a^2$, $\Sigma \equiv r^2 + a^2\cos^2\theta$, and $M$ and $a$ are the mass and specific angular momentum of the background \ac{MBH}, respectively. 
The radial equation for a unit mass test body on an equatorial orbit is:
\bea r^2\frac{dr}{d\tau}=\pm V^{1/2}\,,\label{Oeq:1}\eea
where $V=(Er^2-a x)^2-\Delta(r^2+x^2)\,$, $x = L-aE$, and $d\tau$ is the proper time of the test body. 
The constants 
\bea E=-g_{\mu\nu}(\pd_t)^\mu \frac{dx^\nu}{d\tau}\,,\quad
L=g_{\mu\nu}(\pd_\phi)^\mu \frac{dx^\nu}{d\tau}\,,\eea
are the energy and angular momentum of the test body, respectively. 
In this section, we always assume $L>0$ and let $-M\leq a\leq M$.
So, prograde orbits correspond to $a>0$ and retrograde orbits correspond to $a<0$. 

From \eqref{Oeq:1}, it is clear that the motion is allowed only for $V\geq0\,$.
The potential can be written as $V=r\td{V}\,$, with
\bea\td{V}=(E^2-1)(r-r_a)(r-r_p)(r-r_3)\,,\label{Oeq:2}\eea
where the roots $r_a$, $r_p$ and $r_3$ are fully determined by $M$, $a$, $E$ and $x$.
Because $\td{V}|_{r=0}=2Mx^2\geq0$, the cases with $E\geq1$ cannot have bound orbits, while for the cases with $E<1$ to have bound orbits, the three roots of $\td{V}=0$ must all be positive.
In this paper, we do not expect that an unbound black hole has a significant velocity at spatial infinity.
So, we always assume $E\leq1\,$, where $E<1$ corresponds to bound orbits and $E=1$ corresponds to an unbound parabolic orbit. 

For an elliptical bound orbit, we can assume $r_a\geq r_p\geq r_3>0$ without loss of generality.
In this case, $r_p$ and $r_a$ are the radii of the periastron and apastron of the bound orbit, respectively, and they are related to the semi-latus rectum $p$ and eccentricity $\epsilon$ as
\bea r_p=\frac{p}{1+\epsilon}\,,\quad r_a=\frac{p}{1-\epsilon}\,,\quad 0 \leq \epsilon <1\,.\label{Oeq:rarp}\eea
In this case, 
\bea r_3=\frac{2px^2}{p^2-x^2(1-\epsilon^2)}\,.\eea
The specific energy is
\bea E=\Big\{1-\frac{M(1-\epsilon^2)}{p}\Big[1-\frac{x^2}{p^2}(1-\epsilon^2)\Big]\Big\}^{1/2}\,.\label{Oeq:E}\eea 

\begin{figure}[t!]
\centering
\includegraphics[width=1\linewidth]{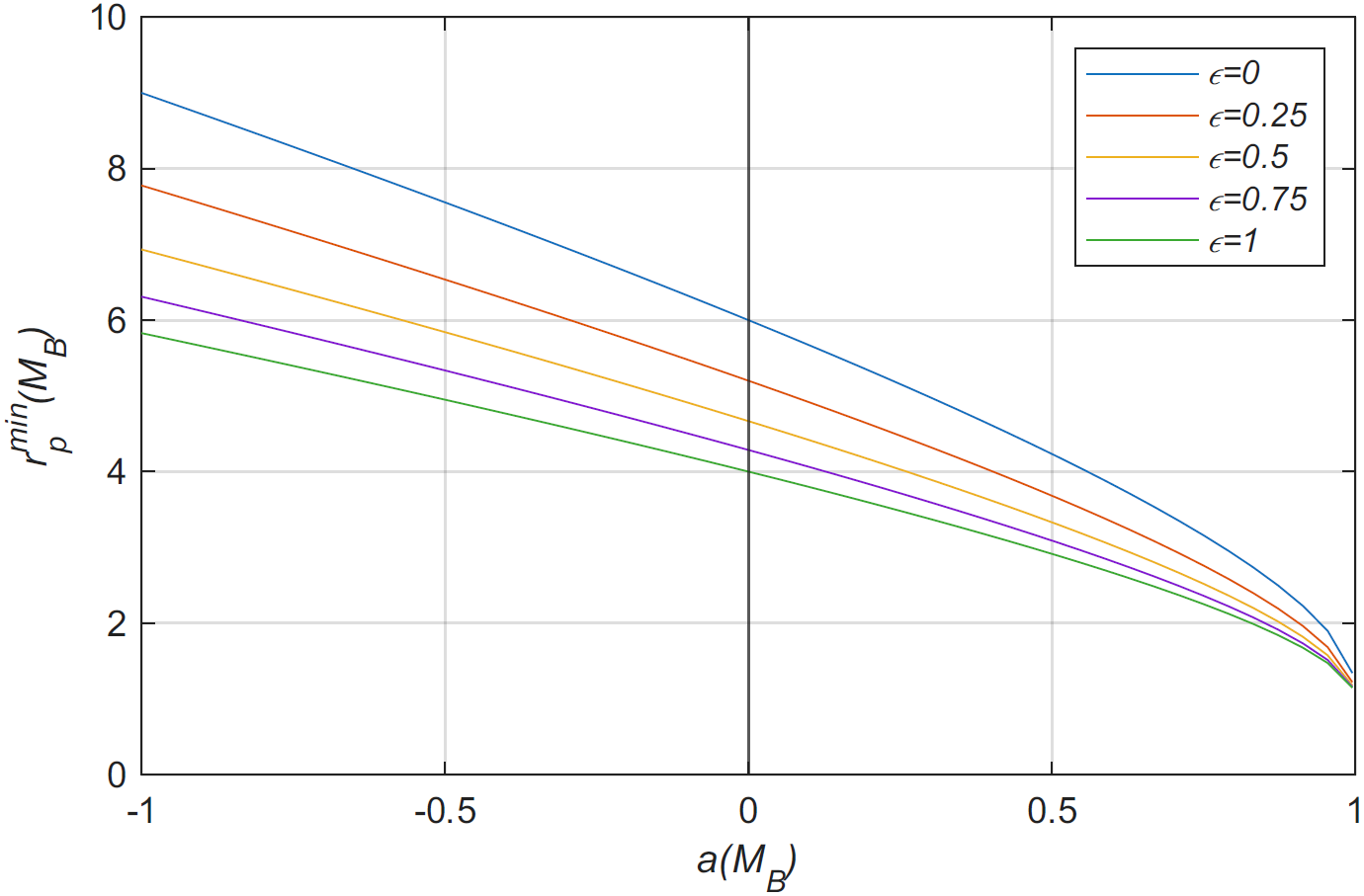}
\caption{The closest periastron for marginally stable elliptical orbits, varing with eccentricity and the specific angular momentum of the background \ac{MBH}.
Note $a<0$ corresponds to retrograde orbits and $a>0$ corresponds to prograde orbits.}
\label{fig:rpmin1}
\end{figure}

In order for the small black holes to have a large initial boost, one would like to minimize $r_p$ as much as possible.
For each fixed value of $r_3$, $r_p>r_3$ corresponds to a stable bound orbit, while $r_p=r_3$ corresponds to a marginally stable orbit \cite{Glampedakis:2002ya}.
So, the minimal value allowed for $r_p$ is achieved when $r_p=r_3\,$, leading to 
\bea r_p^{\rm min}=x\sqrt{\frac{3-\epsilon}{1+\epsilon}} =\frac{p}{1+\epsilon}\,,\eea
where $x$ and $p$ are implicitly determined by $a$ and $\epsilon$, through (\ref{Oeq:E}) and
\bea\frac{(a+Ex)^2}{1-E^2}=\frac{8x^2}{1-\epsilon^2}\,.\eea 
The result is plotted in FIG. \ref{fig:rpmin1}. 

One can see that for orbits with fixed eccentricity, the periastron of a retrograde orbit is always higher than that of a prograde orbit.
So, the lowest collision location in the \ac{MBH} background is determined by the periastron of the retrograde orbit.
Furthermore, for retrograde orbits with fixed eccentricity, the periastron always decreases with the magnitude of the spin of \ac{MBH}, reaching the minimum radius at $a=0$. 
So, in order to maximize the initial boost for the collision, the best scenario is when the background \ac{MBH} is not rotating.

For $a=0$, we can note two extreme cases:
\begin{itemize}
\item $\epsilon=0$: This corresponds to two black holes colliding in a circular orbit on the static \ac{MBH} background. For this case, $r_p^{\rm min}=6M\,$.
\item $\epsilon=1$: This corresponds to two black holes colliding in an extremely eccentric orbit on the static \ac{MBH} background. For this case, $r_p^{\rm min}=4M\,$.
\end{itemize}
For the second case, it appears unnatural for two black holes to be simultaneously placed in the same extremely eccentric orbit.
So we turn to an alternative scenario: the collision of two black holes in the vicinity of \ac{MBH} by falling from infinity.

\begin{figure}[t!]
\centering
\includegraphics[width=1\linewidth]{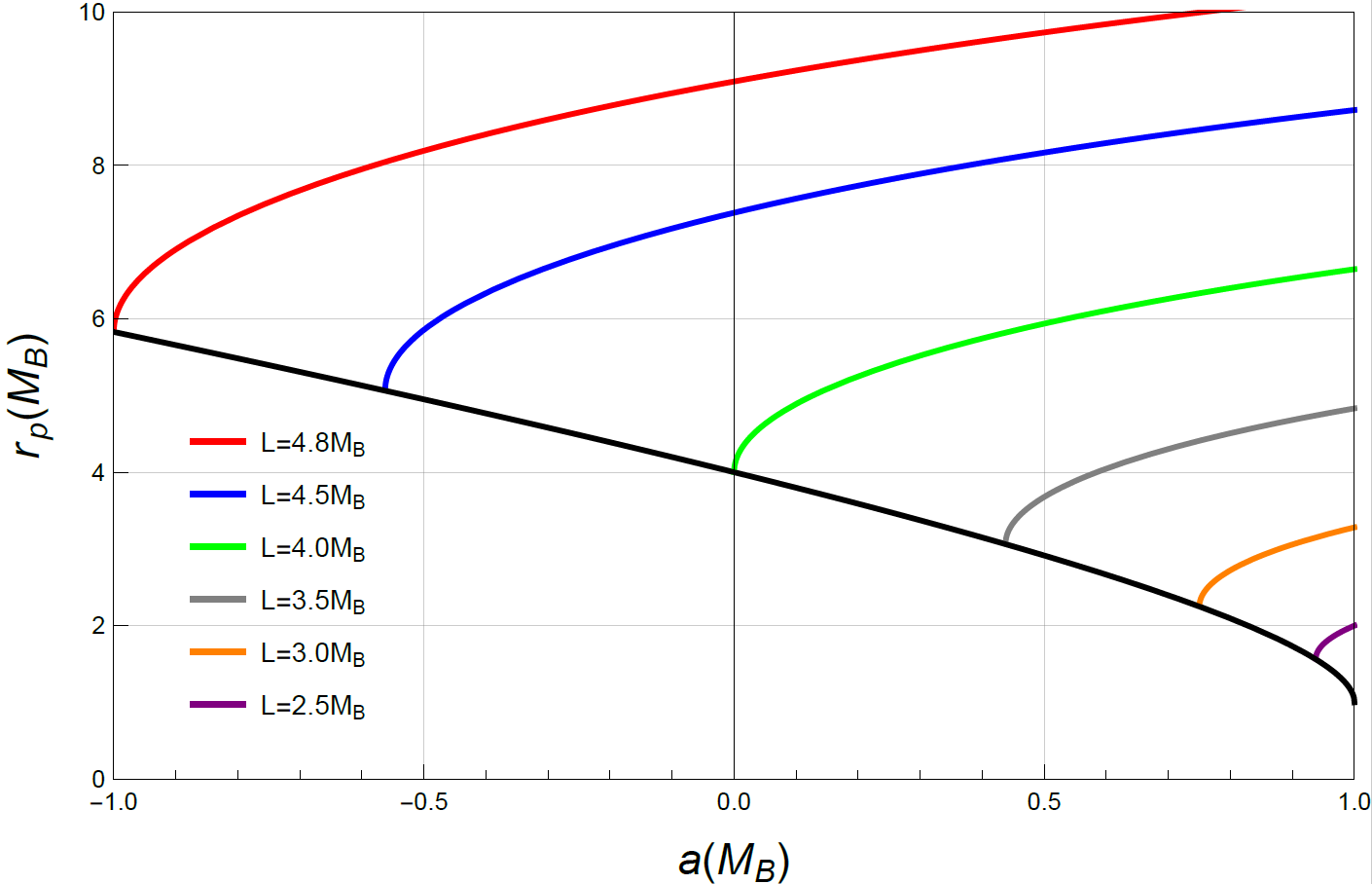}
\caption{The periastron radius of parabolic orbits around Kerr black holes. 
The black curve denotes the closest periastron for all possible values of $L$, and the other colored curves denote orbits with various fixed values of $L$.
Note $a<0$ corresponds to retrograde trajectories and $a>0$ corresponds to prograde trajectories.}
\label{fig:rpmin2}
\end{figure}

The latter case corresponds to an unbound parabolic trajectory, for which we have $E =1$. 
Restoring the rotation of the background \ac{MBH}, the turning point of the trajectory satisfies the equation
\bea 2Mr^2-(x^2+a^2+2ax)r+2Mx^2=0\,.\label{Oeq:4}\eea
Trajectories with the closest periastron are also marginally stable. 
The two repeated roots of \eqref{Oeq:4} are
\bea r_p^{\rm min} =x = 2 M -a + 2 \sqrt{M^2-a M}\,.\eea
The result is illustrated in FIG. \ref{fig:rpmin2}. 

One can see that, as for bound orbits, retrograde and prograde trajectories also have different values for $r_p^{\rm min}$.
The lowest collision location in the \ac{MBH} background is again determined by the retrograde trajectories, and for which the lowest periastron is achieved when $a=0$.
In this case, the two black holes that fall from infinity will have the opportunity to collide at $r_p^{\rm min}=4M$, which is the same as the bound orbit case with $a\rightarrow0$ and $\epsilon\rightarrow1\,$.
For these reasons, we will focus on the non-rotating \ac{MBH} background from now on.
As a result, we will not consider cases that have a nearly infinite collision energy, which relies on having a nearly extremal \ac{MBH} background \cite{Banados:2009pr,Jacobson:2009zg,Berti:2009bk,Berti:2014lva,Harada:2014vka,Mummery:2025cog}.
We hope to investigate these cases in the future.

To quantify the boost, we need to find the momentum of each of the small black holes observed by a local static observer at the collision location. 
This can be done by splitting the four momentum of each of the small black holes as follows.
\bea P^\mu\equiv m_0\frac{dx^\mu}{d\tau}={\cal E}T^\mu+P_0 S^\mu\,,\eea
where $m_0$ is the rest mass of the black hole, $T^\mu$ is the unit time-like vector of the observer, $S^\mu$ is a unit space-like vector orthogonal to $T^\mu$, and ${\cal E}$ and $P_0$ are the energy and momentum of the black hole as seen by the local observer, respectively.

For the cases we are focusing on,
\bea T^\mu&=&\frac1{\sqrt{1-2M/r_p}}(\pd_t)^\mu\,,\nn\\
\Longrightarrow\quad {\cal E}&=&\frac{m_0E}{\sqrt{1-2M/r_p}}\,.\eea
From ${\cal E}^2=m_0^2+P_0^2$, one can obtain the boost ratio:
\bea \frac{P_0}{m_0}&=&\sqrt{\frac{E^2}{1-2M/r_p} -1}\,.\label{eq:P0}\eea
For the marginally stable orbits needed to reach $r_p^{\rm min}\,$, we have $E=\sqrt{8/(9-\epsilon^2)}$ .

Let us consider in more detail the special cases of {\it symmetric collision}, in which the two small black holes have equal masses, zero spins, moving oppositely in a same bound orbit or an unbound parabolic trajectory in the non-rotating \ac{MBH} background:
\begin{itemize}
\item If the two black holes are on a circular bound orbit, then the maximal possible boost ratio is $P_0/m_0\approx0.58$, corresponding to a speed $v\approx0.5$, when the orbit has the minimal possible radius $r=6M\,$, which is the \ac{ISCO} of the Schwarzschild \ac{MBH}.
\item If the two black holes are on an eccentric orbit or an unbound parabolic trajectory and collide at the periastron, then the maximal possible boost ratio is $P_0/m_0=1$, when the lowest periastron is reach at $r_p=4M\,$, corresponding to a speed $v\approx0.7$. 
\end{itemize}
So, limited to the symmetric collision described above, the maximal possible boost ratio from the first stage is $P_0/m_0=1$.

The acceleration of the first stage ends when the two small black holes reach the region where their mutual attraction starts to dominate their acceleration along the collision axis.
Because the \ac{MBH} is assumed to be many orders more massive than the small black holes, one can approximate the background spacetime as flat for the second stage of acceleration.
We further assume that the two small black holes are approximately infinitely apart when they first enter the mutual attraction region.

Because of the large speed of the small black holes at the beginning of the second stage, there is no reliable analytical method to estimate the additional boost that the small black holes can acquire. 
For example, if one were to use the \ac{PN} theory, then the acceleration of each small black hole along the collision axis is given by \cite{Blanchet:2001yf,Blanchet:2002mb},
\bea \frac{dv}{dt}&=&-\frac{m_0}{4r^2}\Big\{1+\frac{1}{c^2}\Big(-\frac{9m_0}{2r} -\frac{17v^2}{2}\Big)\nn\\
&&+\frac{1}{c^4}\Big(\frac{231m_0^2}{16r^2} -\frac{10m_0v^2}{r} -\frac{105v^4}{8}\Big)\nn\\
&&+\frac{1}{c^5}\Big(-\frac{32m_0^2v}{15r^2}-\frac{32m_0v^3}{5r}\Big)\nn\\
&&+\cO\Big(\frac{1}{c^6}\Big)\Big\}\,.\label{eq:PN}\eea
Here, the coordinates are chosen so that the two black holes, each of which has a rest mass $m_0$, move toward each other along the $\hat{z}$ axis, with their locations given by $\vec{r}_\pm(t)=\pm r(t)\hat{z}$, and $v(t)=dr(t)/dt\,$.
The speed of light, $c$, is explicitly shown to indicate the corresponding \ac{PN} order for each term in the equation.
It is clear that in order for the 1PN terms (those proportional to $\frac{1}{c^2}$) to be smaller than the Newtonian term, one must have $v<<\sqrt{2/17}\approx0.34$, which is significantly smaller than the speed that the small black holes can have at the beginning of the second stage.
For this reason, we will not attempt to do any precise calculation for the second stage but instead, use the Newtonian potential as an order of magnitude indicator showing how much gravitational potential energy can be converted into kinetic energy during the second stage of acceleration.

The second stage ends when an encompassing apparent horizon forms around the two small black holes.
For different momentum, the encompassing horizon forms when the two small black holes are at different proper distances apart,
\bea \frac{L_P}{2m_0}\approx1.88+0.48\Big(\frac{P}{2m_0}\Big)+0.27\Big(\frac{P}{2m_0}\Big)^2\,,\label{eq:LPP}\eea
where $L_P$ is the proper distance between the two small black holes, both having momenta $P$, at the formation of an encompacing apparent horizon.
The above formula has been fitted from Fig. 1 of \cite{Abrahams:1994qu}, where $P/(2m_0)\in[0,1.8]\,$.
Using the final relativistic mass $(m_0^2+P^2)^{1/2}$ in the Newtonian potential, the final momentum $P$ can be approximately related the momentum $P_0$ at the beginning of the second stage as
\bea\sqrt{m_0^2+P^2}\approx\sqrt{m_0^2+P_0^2}+\frac{m_0^2+P^2}{2L_P}\,.\eea
Combined with \eqref{eq:LPP}, one can find the following approximate relation,
\bea \frac{P}{m_0}\approx0.55+0.41\Big(\frac{P_0}{m_0}\Big)+0.46\Big(\frac{P_0}{m_0}\Big)^2\,.\eea
In this case, $P/m_0\approx0.94$ for $P_0/m_0=0.58$, and $P/m_0\approx1.42$ for $P_0/m_0=1$.

In the rest of the paper, we will focus on the case of the unbound parabolic trajectory for which the boost from the first stage of acceleration is given by \eqref{eq:P0} with $E=1$, and the final boost after the second stage of acceleration is roughly in the range $P/m_0\in[0.55,1.4]$.

\section{Waveform}
\label{sec:wav}

\begin{figure}[h!]
\centering
\includegraphics[width=1\linewidth]{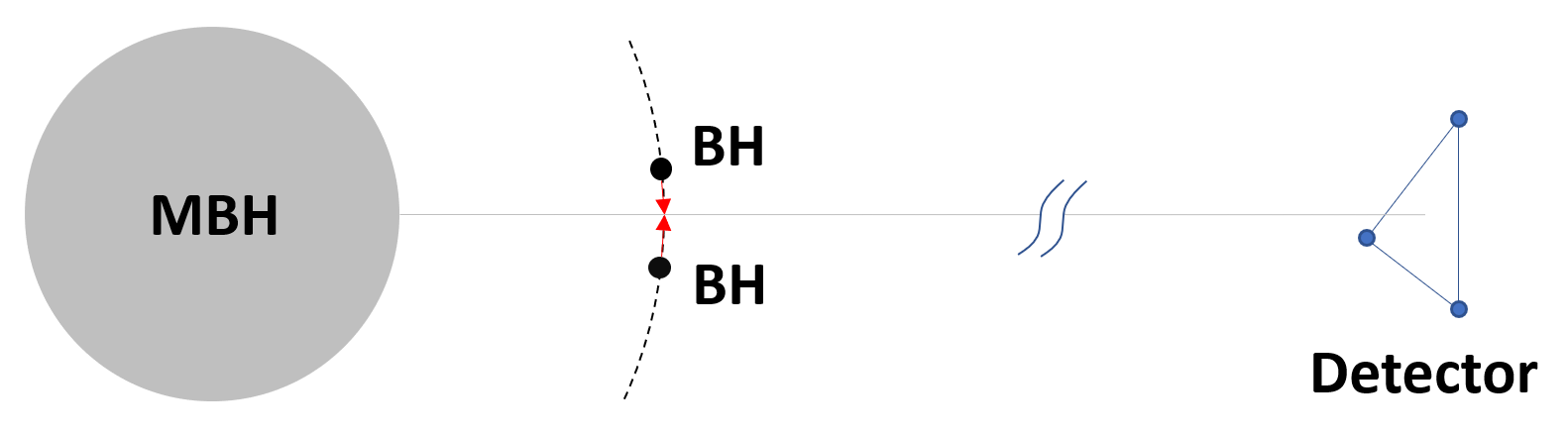}
\caption{An illustration of the detection scenario considered in this paper.}
\label{fig:scheme}
\end{figure}

Due to the symmetry of the collision, the amplitude of \ac{GW} is zero along the axis of collision and is maximal on the transverse plane that intersects the axis at the collision point.  
For simplicity, we assume that the collision is located on the line connecting the center of \ac{MBH} and the detector, so that the detector is in the plane where the amplitude of \ac{GW} is maximal (See Fig. \ref{fig:scheme}).

Apart from giving the small black holes the needed initial boost, the background \ac{MBH} also affects the propagation of the \acp{GW} resulting from the collision.
Since the background \ac{MBH} is assumed to be many orders of magnitude more massive than the small black holes, the geometric optics approximation can be used for the radiated \acp{GW} \cite{Isaacson:1968hbi}.
In the leading order, the amplitude of \ac{GW} is not affected by the \ac{MBH}, but there is a gravitational redshift to the \ac{GW} frequency,
\bea f&=&f_0\Big(1-\frac{2M}{r_p}\Big)^{1/2}(1+z)^{-1}\nn\\
&=&\frac{f_0E}{(1+z)\sqrt{1+(P/m_0)^2}}\,,\label{redshift.freq}\eea
where the \ac{MBH} is assumed to be located at redshift $z$, $f_0$ is the frequency observed by a local static observer at the collision point, and $f$ is the \ac{GW} frequency observed by a distant observer at redshift $0$.

With this in mind, we take the following steps to construct the waveform:
\begin{itemize}
\item Step 1: The initial waveform of the collision is calculated at the location of the collision. 
Since the size of the collision region is many orders of magnitude smaller than the radius of the \ac{MBH}, we approximate the background as a flat spacetime for this step.
\item Step 2: The waveform is redshiftted as in \eqref{redshift.freq} to give raise to the signal that reaches the detector. 
\end{itemize}
In the rest of the section, we explain in more detail how step 1 is achieved.

To obtain the waveforms at the collision location, we use the \ac{CLAP} originally proposed to describe the head-on collision of static black holes that start from a close distance apart \cite{Price:1994pm}.
The method is based on the observation that, after a common horizon forms around the small black holes, the spacetime outside the horizon evolves as a perturbed single black hole.
The method has been shown to make predictions that match the numerical relativity results quite well \cite{Anninos:1995vf,Abrahams:1995gn,Abrahams:1995wd}.
It has also been generalized to describe the collision of black holes that have initial momenta \cite{Abrahams:1994qu,Baker:1996bt,Nicasio:1998aj}.
However, when comparing the radiated energy predicted by \ac{CLAP} (e.g. \cite{Nicasio:1998aj}) with that by more recent numerical relativity simulations (e.g. \cite{Sperhake:2008ga}), we note that there is a significant difference. 
For example, when $P/m_0=1$, corresponding to $P/M_{\rm ADM}\approx0.55$ in FIG.1 of \cite{Nicasio:1998aj} and $\beta\approx0.7$ in FIG. 3 of \cite{Sperhake:2008ga}, the radiated energy predicted by \cite{Sperhake:2008ga} is approximately 5 times that of \cite{Nicasio:1998aj}.
If both calculations are correct, it indicates that a significant part of the radiated energy, some of which is possibly generated before the formation of the common horizon, is not taken into account by \ac{CLAP}.
Therefore, when using the waveform calculated using \ac{CLAP} to assess the detectability of \acp{RBHC}, the results should be considered pessimistic.

In the following, we describe the detailed steps of \ac{CLAP} used in this paper.
There is not much new here, but it could be helpful for the reader to clearly see how various ingredients of \ac{CLAP} are grouped together.
We mainly follow the presentation of \cite{Martel:2005ir,Ruiz:2007yx} for the Zerilli equation, 
\cite{Misner:1960,Price:1994pm} for the Misner initial data, and 
\cite{Baker:1996bt,Brandt:1997tf,Nicasio:1998aj} for adding the correction of initial boost.

\subsection{The Zerilli equation}

After the formation of a common horizon, the evolution of spacetime outside the horizon is described by the perturbation of a single black hole, which in this paper is a Schwarzschild black hole with the metric,
\bea g_{\mu\nu}dx^{\mu}dx^{\nu}=-X(r)dt^{2}+\frac{1}{X(r)}dr^{2}+r^{2}d\Omega^{2} \,,\eea
where $X(r) = 1-2m/r$, and $m$ is the ADM mass of the black hole. 
The metric can be split as
\bea ds^2=g_{ab}dx^adx^b+r^2\Omega_{AB}dx^Adx^B\,,\eea
where $x^a=(t,r)$ and $x^A=(\theta, \phi)\,$. 
$\Omega_{AB}$ is the metric of a unit two-sphere. 
The covariant derivatives associated with $g_{ab}$ and $\Omega_{AB}$ will be denoted as $\nabla_a$ and $\nabla_A$, respectively. 

Due to the symmetry of the collision, we only need to consider the even-parity perturbation of the metric, which is given by
\bea h_{ab}&=&\sum_{lm}h_{ab}^{lm}Y^{lm}\,,\quad 
h_{aB}=\sum_{lm}j_{a}^{lm}Y_{B}^{lm}\,,\nn\\
h_{AB}&=&r^{2}\sum_{lm}\Big(K^{lm}\Omega_{AB}Y^{lm}+G^{lm}Y_{AB}^{lm}\Big)\,.\label{eq:S3}\eea
Here, $h_{ab}^{lm}$, $j_{a}^{lm}$, $K^{lm}$, and $G^{lm}$ are functions of $x^a$. 
$Y^{lm}(\theta,\phi)$ are sphereical harmonics, from which one can construct the even-parity vector harmonics, $Y^{lm}_A=\nabla_AY^{lm}$, and the even-parity tensor harmonics,
\bea  Y^{lm}_{AB}=\Big[\nabla_A \nabla_B + \frac{1}{2}l(l+1)\Omega_{AB}\Big]Y^{lm} \,.\eea
The perturbation is governed by the Zerilli equation,
\bea  \frac{\pd^2 \Psi^{lm}}{\pd r_*^2} - \frac{\pd^2 \Psi^{lm}}{\pd t^2}-V_{eff} \Psi^{lm}=0 \,,\label{eq:zerilli}\eea 
where \cite{Martel:2005ir,Ruiz:2007yx}
\bea \Psi^{lm}(t,r)&=&\frac{2r}{l(l+1)}\Big[\tilde{K}^{lm}+\frac{1}{n+3m/r}\nn\\
&&\times\Big(r^{a}r^{b}\tilde{h}_{ab}^{lm}-rr^{a}\nabla_{a}\tilde{K}^{lm}\Big)\Big]  \label{eq:Z-M} \,,\\
V_{eff}(r)&=&\frac{2}{3r^2}\Big(1-\frac{2m}{r}\Big)\Big(n+\frac{3m}{r}\Big)\nn\\
&&\times\Big[1+\frac{n^2(3+2n)}{(n+3m/r)^3}\Big] \,. \eea 
 Here, $r^a =g^{ab} \pd_b r$ and $n = (l-1)(l+2)/2$, $\tilde{K}^{lm}$ and $\tilde{h}_{ab}^{lm}$ are gauge invariant combinations of $h_{ab}^{lm}$, $j_{a}^{lm}$, $K^{lm}$, and $G^{lm}$, see \cite{Martel:2005ir} for detailed definitions. 

Once $\Psi^{lm}$ is found by integrating \eqref{eq:zerilli}, the relevant part of the waveform in the transverse and traceless gauge can be obtained as \cite{Ruiz:2007yx},
\bea (h^{+})^{lm} =\frac{\Psi^{lm}|_{r\rightarrow\infty}}{D_L}\left[\frac{\pd^2}{\pd\theta^2}+\frac{1}{2}l(l+1)\right]Y^{lm}\,,\label{def:hplus}\eea
where $D_L$ is the luminosity distance of the \ac{MBH}.
After being redshifted according to \eqref{redshift.freq}, the above formula gives the waveform that will be used to assess the capabilities of different detectors.

\subsection{The Misner initial data}

To numerically integrate \eqref{eq:zerilli} for a general perturbation, one needs to set the initial value for $\Psi^{lm}(t=0,r)$ and $\dot\Psi^{lm}(t=0,r)$, where the dot means a time derivative.

To see how the initial values are determined, note that under the $3+1$ decomposition with Gaussian normal coordinates, the four dimensional spacetime metric can be written as (see, e.g., \cite{Gourgoulhon:2007ue}),
\bea \mathrm{d}s^{2}=-\mathrm{d}T^{2} +\gamma_{ij} \mathrm{d}x^{i} \mathrm{d}x^{j}\,,\eea
where $i,j = 1,2,3$.
On any hypersurface with constant $T$, the three dimensinal metric $\gamma_{ij}$ obeys the constraint equations, 
\bea
^3R+K^2-K_{ij}K^{ij}&=&0 \label{eq:initial1}\\
D_j(K^{ij}-\gamma^{ij}K)&=&0 \label{eq:initial2}
\eea
where $^3R$ and $K_{ij}$ are the scalar curvature and the external curvature tensor of the hypersurface, respectively, $K = K_{ij}\gamma^{ij}$, and $D_i$ is the covariant derivative.
Assuming that the $T=0$ hypersurface is conformally flat, with $\gamma_{ij} = \Phi^4 \hat{\gamma}_{ij}\,$, 
$K_{ij} = \Phi^{-2}\hat{K}_{ij}\,$, then \eqref{eq:initial1} and \eqref{eq:initial2} can be written as
\bea \hat{D}^i\hat{K}_{ij} &=& 0\,,\label{eq:Khat1}\\
\hat{D}^i\hat{D}_i\Phi &=& -\frac{1}{8}\Phi^{-7}\hat{K}^{ij}\hat{K}_{ij}\,, \label{eq:Khat2}\eea
where 
$\Phi^4$ is the conformal factor, 
$\hat{\gamma}_{ij}$ is the flat metric, 
and $\hat{D}^i$ is the flat space covariant derivative. 

The initial formulation of \ac{CLAP} \cite{Price:1994pm} uses the Misner initial data \cite{Misner:1960}, which is a solution of the constraint equations \eqref{eq:initial1}, \eqref{eq:initial2}, \eqref{eq:Khat1}, \eqref{eq:Khat2}, with $K_{ij}=\hat{K}_{ij}=0$, and describes two equal mass and initially static black holes,
\bea ds_{\text{Misner}}^{2}&=&\gamma_{ij} \mathrm{d}x^{i} \mathrm{d}x^{j}|_{T=0}\nn\\
&=&a^{2}\varphi_{\mathrm{Misner}}^{4}\left[d\mu^{2}+d\eta^{2}+\sin^{2}\eta d\phi^{2}\right]\,,\eea
where $(\mu,\eta,\phi)$ are bispherical coordinates and
\bea\varphi_{\text{Misner}}=\sum_{n=-\infty}^{n=-\infty}\frac{1}{\sqrt{\cosh(\mu+2n\mu_{0})-\cos\eta}}\,.\eea

The ADM mass of Misner data $m^\prime$ is denfined as
\bea m' &=& 4a\Sigma_1(\mu_0) \,, \eea
where $\Sigma_1(\mu_0) = \sum^\infty_{n=1}(\sinh{n\mu_0})^{-1}$.
The proper distance between the two black holes is \cite{Anninos:1995vf}
\bea L_P=2a\Sigma_2(\mu_0)\,,\eea
where $\Sigma_2(\mu_0) =1+2\mu_0\sum_{n=1}^\infty\frac{n}{\sinh n\mu_0}\,$. 
The rest mass of one of the black hole is \cite{Smarr:1976}
\bea m_0 = \frac{\Sigma_3}{2\Sigma_1} m'\,,\eea
where $\Sigma_3 = \sum_{n=1}^\infty\frac{n}{\sinh{n\mu_0}}$.

In our calculation, we will always truncate the summation in $\Sigma_1(\mu_0)$ and $\Sigma_2(\mu_0)$ to $n=1000$ for both definitions. 
One can check that $L_P/m'$ increases monotonically with $\mu_0$, with $L_P/m'\rightarrow 0$ as $\mu_0 \rightarrow0$, and $L_P/m'\rightarrow\mu_0$ as $\mu_0 \rightarrow+\infty$.

Introducing spherical coordinates through the following coordinate transformation,
\bea R\equiv a\sqrt{\frac{\cosh\mu+\cos\eta}{\cosh\mu-\cos\eta}}\,,\quad
\tan\theta\equiv\frac{\sin\eta}{\sinh\mu}\,,\eea
the Misner metric can be written as
\bea ds_{\text{Misner}}^{2}&=&\Phi^{4}\Big[dR^{2}+R^{2} (d\theta^{2} +\sin^{2}\theta d\phi^{2})\Big] \,. \label{metricprime:Misner}\\
\Phi&=&\left(1+\frac{m^\prime}{2R}\right)\mathcal{F} \,, \label{conformal_factor} \eea
where
\bea&&\mathcal{F}\equiv1+\frac{2}{X_2}\sum_{l=2,4,...}\kappa_{l}(\mu_{0})\left(m^\prime/R\right)^{\l+1}P_{l}(\cos\theta) \,,\nn\\
&&\kappa_l(\mu_0)\equiv\frac{1}{[4\Sigma_1(\mu_0)]^{l+1}} \sum_{n=1}^\infty\frac{(\coth n\mu_0)^l}{\sinh n\mu_0}\,.\eea
Here $X_2=1 +m'/(2R)\,$ and $P_l(\cos\theta)$ is the Legendre functions. 
One can further introduce the Boyer-Lindquist radial coordinate with the transformation,
\bea R=\frac{r^2}{4}\left(1+\sqrt{X_1}\right)^2\,, \label{coordinates_trans}\eea
where $X_1=1-2m'/r\,$.
As a result, 
\bea&&ds^2_{\text{Misner}} = \mathcal{F}^4\Big[\frac{dr^2}{X_1} + r^2 (d\theta^{2} +\sin^{2}\theta d\phi^{2})\Big]\,. \label{metric:Misner}\eea

In the small $\mu_0$ limit, $\kappa_l$ can be simplified as
\bea \kappa_l\approx\frac{\zeta(l+1)}{|4\ln\mu_0|^{l+1}} \,,
\eea 
where $\zeta(s)$ is Riemann's $\zeta$ function. Using $1/|\ln\mu_0|$ as a small expansion parameter, $\mathcal{F}^4$ can be written as
\bea  \mathcal{F}^4\approx 1+\frac{8}{X_2}\sum_{l=2,4,\ldots}^{\infty} \kappa_{l}\Big(\frac{m'}{R}\Big)^{l+1}P_l(\cos\theta) \,.\eea 
With a coordinate transformation $t^a = \sqrt{X_1}T^a$, the complete four dimensional metric takes the form
\bea  ds^2|_{t=0}=-X_1dt^2+\mathcal{F}^4\Big[\frac{dr^2}{X_1}+r^2d\Omega^2\Big] \,,
\eea 
which obeys $\pd_t g^{\prime}_{\mu\nu}|_{t=0}= 0$ .
Compared with \eqref{eq:S3}, the two nonzero terms in \eqref{eq:Z-M} can be found as
\bea&&h_{rr}^{lm}= \frac{16}{X_1X_2}\sqrt{\frac{\pi}{2l+1}}\kappa_l (\frac{m'}{R})^{l+1}\,,\nn\\
&&K^{lm}=\frac{16}{X_2}\sqrt{\frac{\pi}{2l+1}}\kappa_l (\frac{m'}{R})^{l+1}\,.\eea
This determines $\Psi^{lm}|_{t=0}$. 
Together with $\dot\Psi^{lm}|_{t=0}=0$, we have initial data to describe the collision of two initially static black holes. 

It is remarkable that, although the above derivation requires $\mu_0\rightarrow0$, \ac{CLAP} has been shown to agree well with numerical relativity for $\mu_0 < 2$ \cite{Baker:1996bt,Nicasio:1998aj}. 

\subsection{Adding the correction of initial boost}

The correction of a non-zero momentum to the initial data can be added following \cite{Baker:1996bt,Nicasio:1998aj}.
In the flat space with the metric $\hat\gamma_{ij}$ and Cartesian coordinates $\{x',y',z'\}$, a solution of \eqref{eq:Khat1} that describes a black hole with momentum $P$ can be written as \cite{Baker:1996bt}
\bea\hat{K}_{ij}^{\mathrm{one}}(\vec{R}',P)=\frac{3}{2R'^2} \Big[2P_{(i}n_{j)}-(\delta_{ij}-n_{i}n_{j})P^{k}n_{k}\Big]\label{eq:Khat3} \,, \eea
where $R'$ and $\hat{n}^i$ are the length and unit vector of the radial coordinate $\vec{R}'$, and $P^i$ are the components of the constant momentum $P$, with $\hat{D}_iP^i = 0$. 
The indices are raised and lowered using $\hat{\gamma}_{ij}$.
For the initial data describing two black holes boosted toward each other, one can write
\bea \widehat{K}_{ij}^{\mathrm{two}}&=&\widehat{K}_{ij}^{\mathrm{one}} (\vec{R}'^+,P) +\widehat{K}_{ij}^{\mathrm{one}} (\vec{R}'^-,-P)\label{eq:Khat4}\,,\nn\\
\vec{R}'^\pm&=&\{x',y',z'\pm L/2\}\,,\eea
where positive $P$ is assumed to be in the positive $\hat{z}'$ direction, and $L$ is the distance between the centers of the two black holes in the $\hat{\gamma}_{ij}$ space
\bea L = \frac{\coth{\mu_0}}{2\Sigma_1}m^\prime \,.\eea
Keeping only corrections to the linear order in $P$, there is no correction to the right hand side of \eqref{eq:Khat2} or $\Psi^{lm}|_{t=0}$, except that the ADM mass of the whole spacetime should be modified as \cite{Brandt:1997tf},
\bea m&=& m'\Big[1 +\frac{11}{25} \Big(\frac{P}{m'}\Big)^2\Big(\frac{L}{m'}\Big)^2\Big]\,. \label{MMeq:3}\eea
Note that we have only kept the leading correction term in the above formula and found good agreement with the numerical results in \cite{Nicasio:1998aj}. 
All other corrections from the momentum correction will only come through $\dot\Psi^{lm}|_{t=0}$.

From \eqref{eq:Khat3} and \eqref{eq:Khat4}, we can obtain the complete form of the extrinsic curvature $K_{ij}$, written in terms of the Boyer-Lindquist coordinates used in \eqref{metric:Misner}. 
In the process, we assume $R>>L$ and $R>>m$, and the conformal factor in \eqref{conformal_factor} is approximated as $\Phi \approx 1+\frac{m}{2R}$. 
Note we have replaced the ADM mass $m'$ with $m$. 
Using \eqref{eq:Z-M} and the time derivative of the four metric $g_{\mu\nu}$ at $t = 0$,
\bea  \frac{\pd g _{0\nu}}{\pd t}|_{t=0} &=& 0\,, \nn\\
\frac{\pd g _{ij}}{\pd t}|_{t=0} &=& -2\sqrt{X(r)}K_{ij}\,, \label{eq:Khat5}\eea 
where $X(r)=1-2m/r$, we find that
\bea \dot{\Psi}|_{t=0} ^{(2,0)} &=& - \frac{ 16 \sqrt{\frac{\pi }{5}} L P \sqrt{X(r)}}{r^3 (3 m+2 r)} \left(1+\frac{3 m}{4 r}+\frac{5 L^2 m}{56 r^3}\right)\,,\nn\\ 
\dot{\Psi}|_{t=0} ^{(4,0)} &=& -\frac{2 L^3 P \sqrt{\pi X(r)}}{3 r^3 (m+3 r)}\left(1+\frac{3 m}{7 r}+\frac{3 L^2 m}{88 r^3}\right)  \,, \nn \\
\dot{\Psi}|_{t=0} ^{(6,0)} &=& -\sqrt{\frac{\pi }{13}} L^5 P \sqrt{X(r)} \frac{35  m+88 r}{44 r^6 (3  m+20 r)}\,,\eea
where for $\dot{\Psi}|_{t=0} ^{(6,0)}$ we have omitted all terms higher than the order $(L/r)^5$.

\subsection{Result}

\begin{figure}[htbp!]
    \centering
    \includegraphics[width=0.9\linewidth]{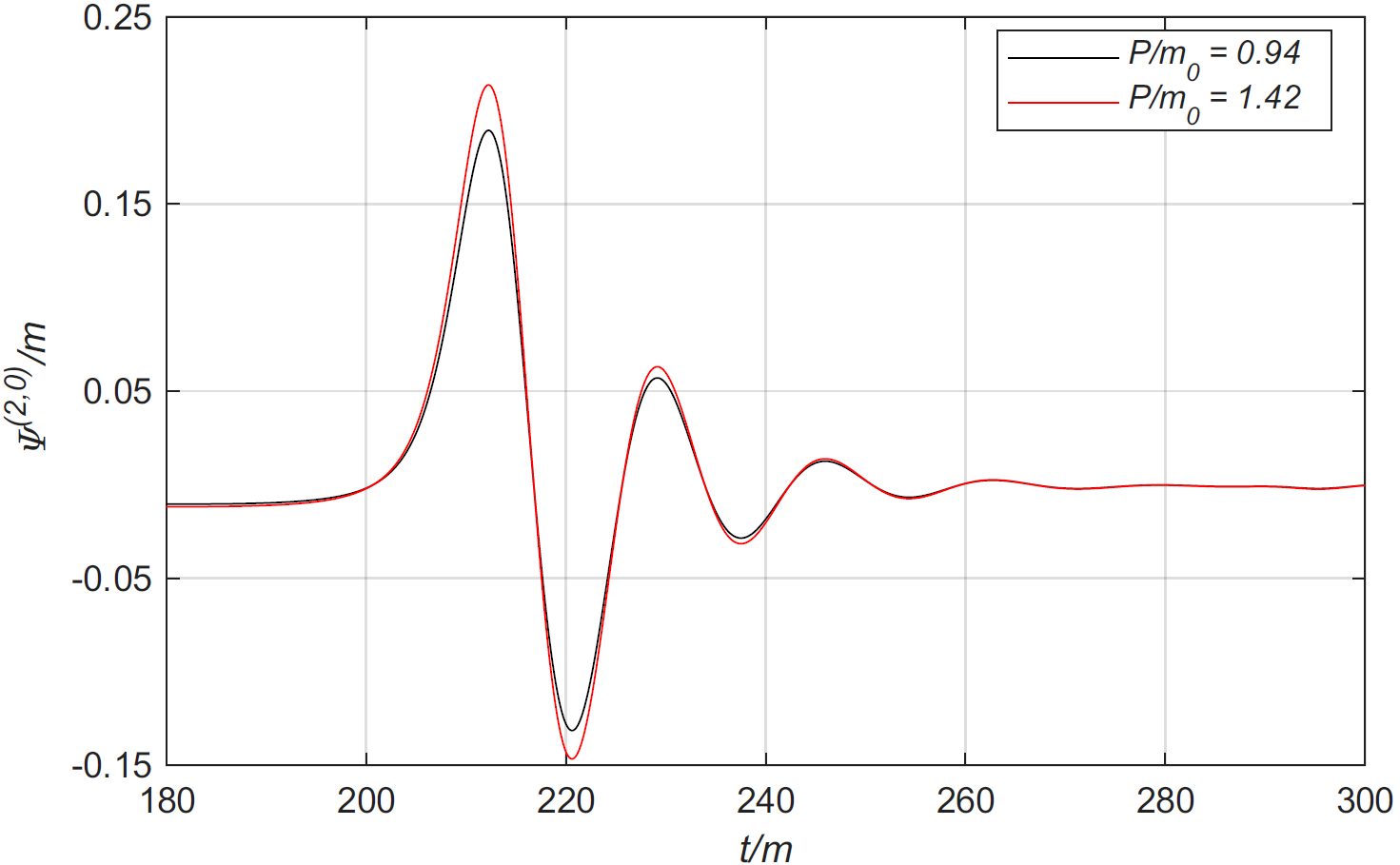}
    \caption{Waveform numerically calculated for two different values of initial boost, as observed by an observer fixed at $r = 200m$.}
    \label{fig:waveform}
\end{figure}

The waveform obtained using the above procedure is illustrated in Fig. \ref{fig:waveform}.
In the figure, $P/m_0=0.94$ corresponds to the case when the two small black holes collide at the \ac{ISCO} of the background Schwarzschild \ac{MBH}, while $P/m_0=1.42$ corresponds to the case when the two small black holes collide at the periastron of the closest scattering orbit.
The two waveforms are similar, but there is a notable difference between their peak amplitudes.

\begin{figure}[htbp!]
    \centering
    \includegraphics[width=0.9\linewidth]{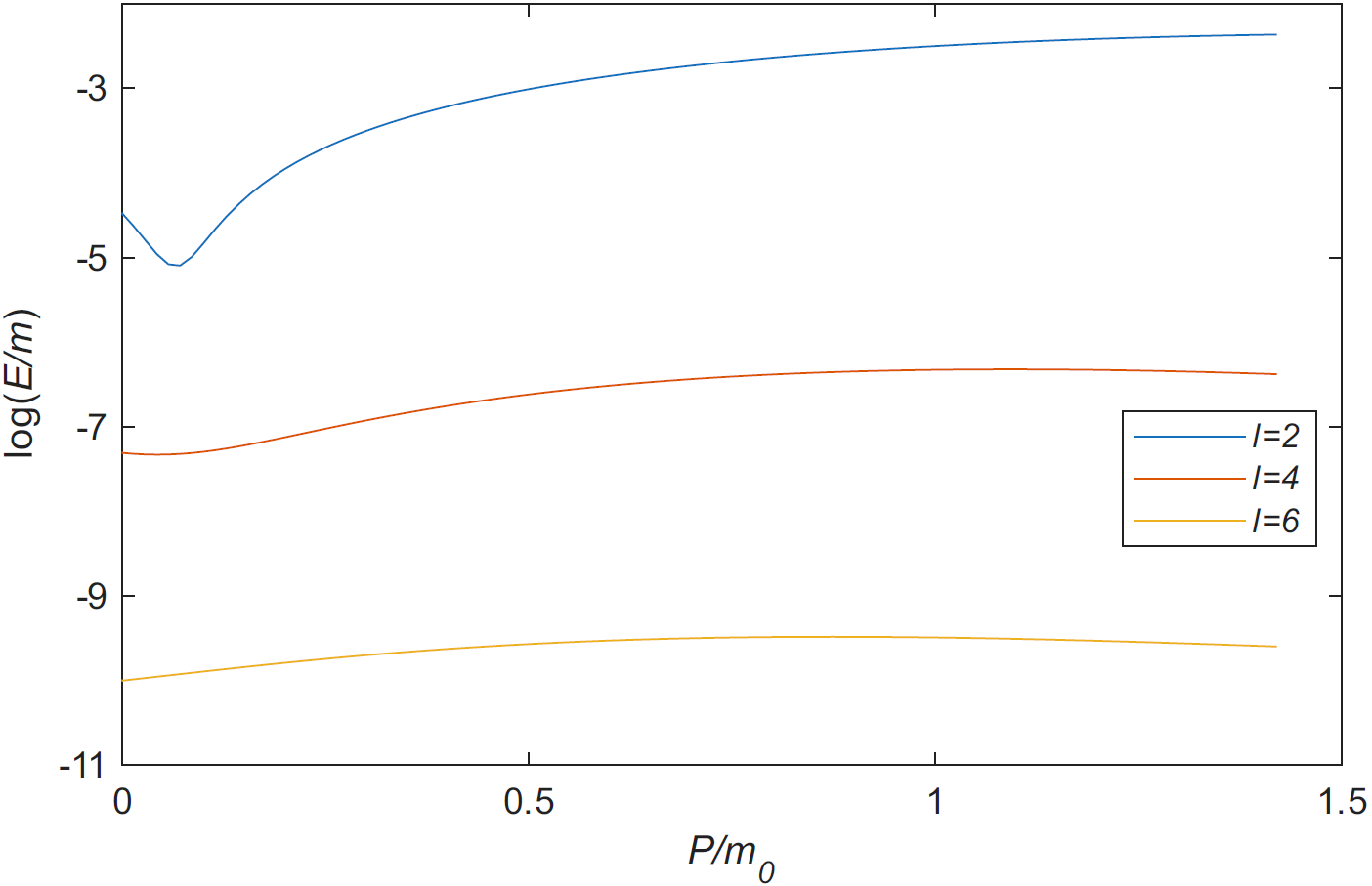}
    \caption{A comparison of the radiated energy of the (2,0), (4,0) and (6,0) modes.}
    \label{fig:multipole}
\end{figure}

A comparison of the radiated energy of the (2,0), (4,0) and (6,0) modes is given in Fig. \ref{fig:multipole}.
Unfortunately, the result for the (4,0) and (6,0) modes appears to be ill behaved for large values of $P/m_0$.
The line for the (4,0) mode starts to bend down for $P/m_0\gtrsim1$, and that for the (6,0) mode starts to bend down for $P/m_0\gtrsim0.75$.
This is counterintuitive, because we expect the contribution of higher modes to become more and more significant as the value of $P/m_0$ increases.
The cause of the problem could be twofold.
Firstly, the numerical result of \cite{Nicasio:1998aj} used to verify \eqref{MMeq:3} only contains the contribution of the (2,0) mode, so the contribution of the higher modes has not been adequately taken into account in \eqref{MMeq:3}.
Secondly, the present calculation only considers the linear order contribution of the initial momentum correction to \eqref{eq:Khat1} and \eqref{eq:Khat2}, which is not adequate when the initial momentum becomes large.
Despite this problem, one can still see that for $P/m_0\lesssim1$ the radiated energy of the (6,0) mode is more than 2 orders of magnitude smaller than that of the (4,0) mode, while the latter is more than 2 orders of magnitude smaller than that of the (2,0) mode.
It will be interesting to see if more realistic collision scenarios, such as when the two small black holes have unequal masses, non-zero spins, and a non-zero impact parameter, will increase the relative contribution of higher modes.

\section{Detection capability}
\label{sec:cap}

In this section, we study the capability of several current and future detectors in detecting \ac{RBHC} signals, including TianQin \cite{Luo:2025ewp}, 
LISA \cite{Robson:2018ifk} (with a similar capability of Taiji \cite{Hu:2017mde}), 
\ac{CE} \cite{cosmicexplorer_sensitivity}, 
\ac{ET} \cite{ET_sensitivities}, 
and LIGO \cite{LIGOT2200043}.
For LIGO, we will always assume the O4 sensitivity.

Before delve into any further details, we note that there is a theoretical upper bound on the masses of \acp{MBH} grown through the luminous accretion of gas \cite{2016MNRAS.456L.109K}.
If other non–luminous means such as black hole merger cannot significantly alter such a bound, we expect all the \acp{MBH} in the Universe to be no more massive than the order $\cO(10^{11})\mSun$.
Consequently, if we want the small black holes to be 5 to 6 orders less massive than the background \ac{MBH}, the masses of the small black holes can at most be of the order $\cO(10^{6})\mSun$.
In the following calculations, however, we will keep the possibility open for having more massive \acp{MBH} in the Universe and let the mass range of small black holes be limited only by the capabilities of detectors.

Firstly, we look at the \ac{SNR} and the detection horizon.
The expected \ac{SNR} can be estimated as
\bea \mathrm{SNR}(h)=\sqrt{(h|h)}\,,\eea 
where the inner product is defined as
\bea (p|q)=2\int_{f_{\mathrm{low}}}^{f_{\mathrm{high}}}\frac{p^*(f)q(f) +p(f)q^*(f)}{S(f)}df\,.\eea
Here $S(f)$ is given by the sensitivity of the relevant detector, which we obtain from the references cited above.

\begin{figure}[htbp!]
    \centering
    \includegraphics[width=1\linewidth]{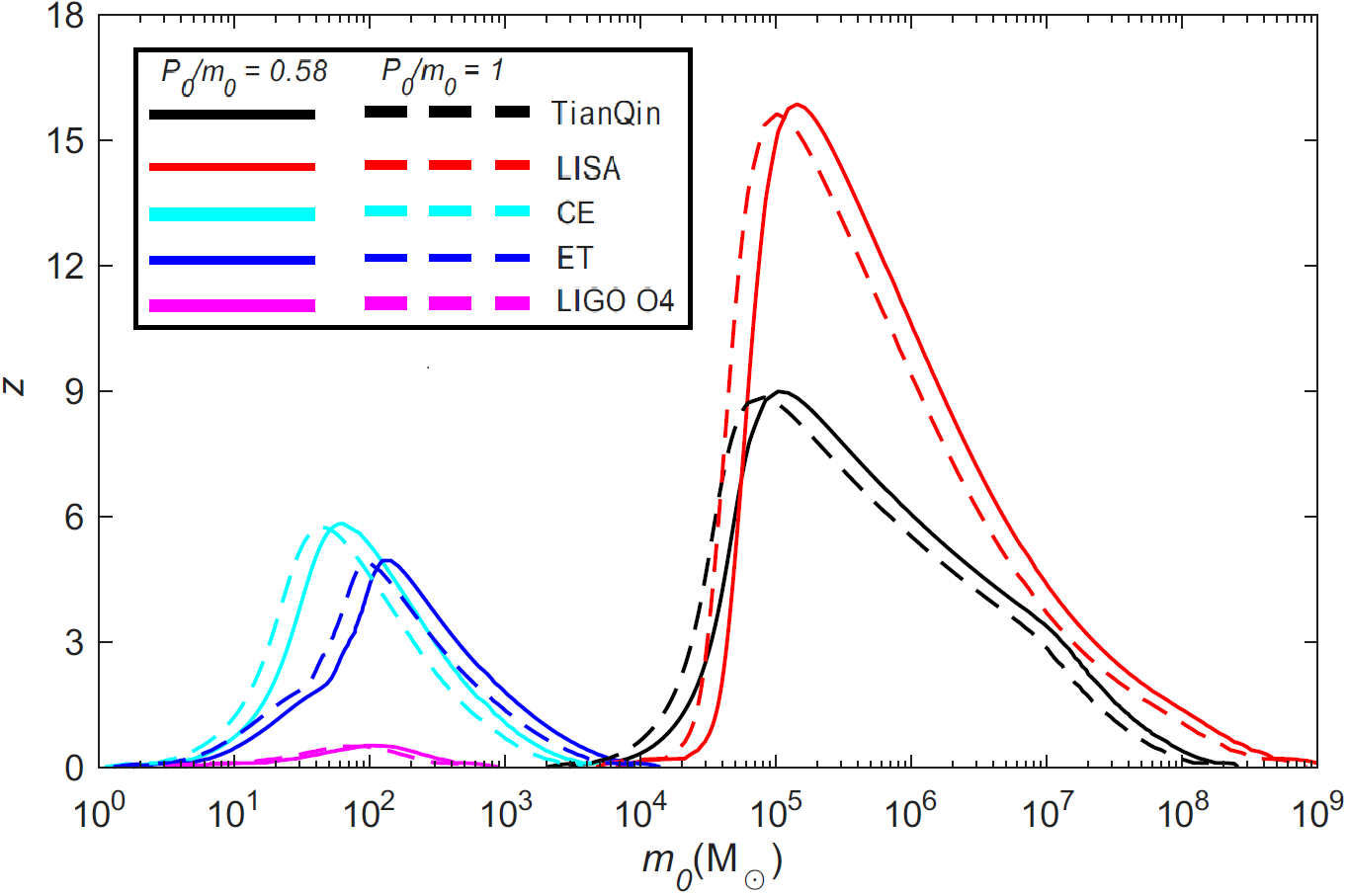}
    \caption{Detection horizon for different detectors in detecting RBHCs, where $\theta = \pi/2$.}
    \label{fig:horizon}
\end{figure}

In Fig. \ref{fig:horizon}, we plot the detection horizon of different detectors assuming the threshold \ac{SNR}$_{\rm thr}=8$ for RBHC at ISCO and closest parabolic orbit separately, where the corresponding momentum after first stage of acceleration is $P_0/m_0 = 0.58$ and $P_0/m_0 = 1$.
One can see that, with an ideal source, TianQin can detect \ac{RBHC} at a redshift of about 9, LISA to about 16, CE to about 6, ET to about 5, and LIGO to about 0.5.
Noteably, both TianQin and LISA reach their farthest detection distances with sources having masses near $1.5\times10^5\mSun$, indicating that there is a physical chance for space-based detectors to detect \acp{RBHC} even if the masses of the most massive \acp{MBH} are bounded at $\cO(10^{11})\mSun$.

\begin{figure}[htbp!]
    \centering
    \includegraphics[width=1\linewidth]{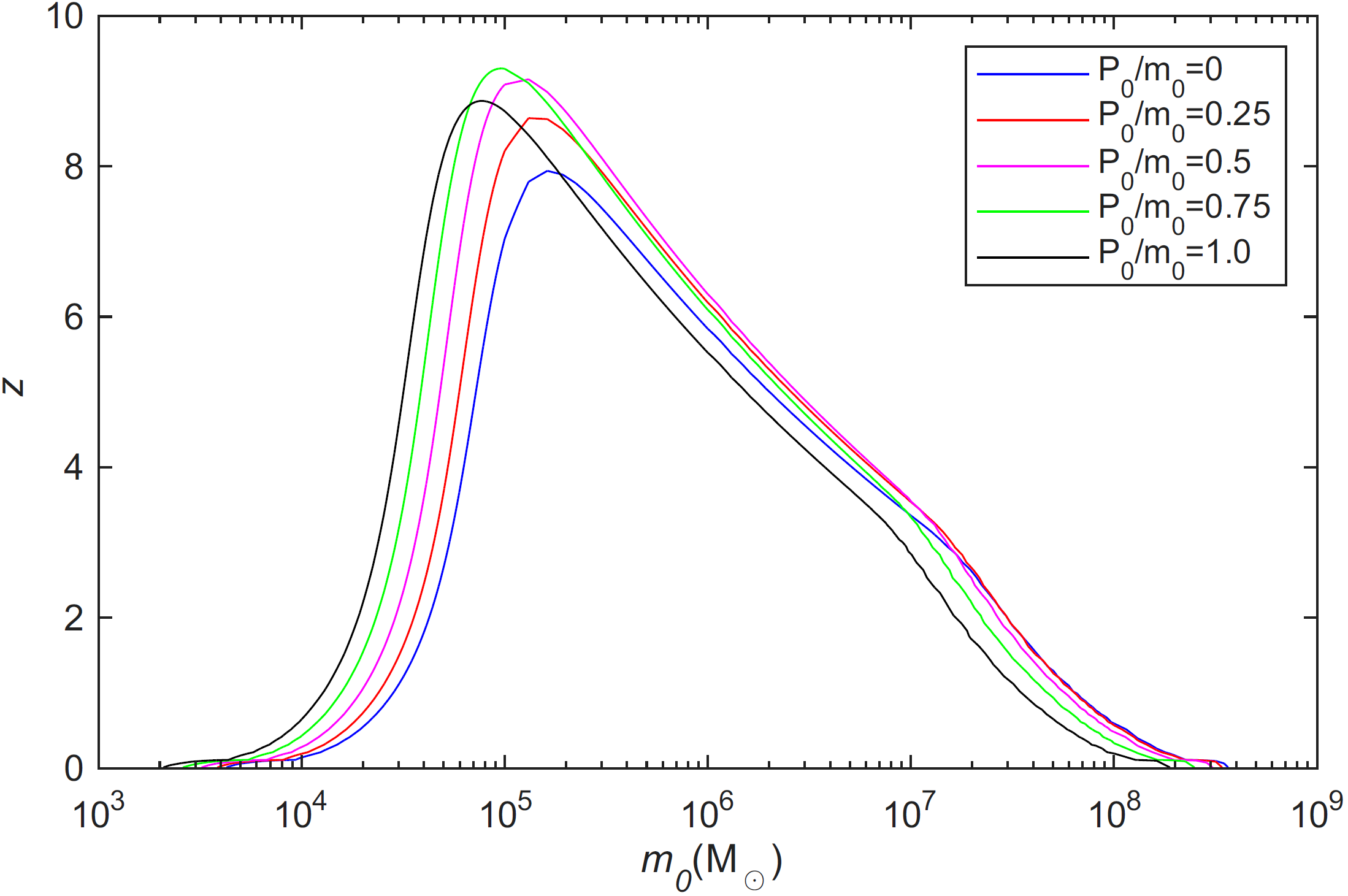}
    \caption{The detection horizon of TianQin for different values of $P_0/m_0$ for parabolic orbits.}
    \label{fig:horizon2}
\end{figure}

In Fig. \ref{fig:horizon}, the detection horizons for signals with $P_0/m_0=0.58$ are greater than those for signals with $P_0/m_0=1$. 
This is unexpected.
To see the dependence more clearly, we plot in Fig. \ref{fig:horizon2} the detection horizon of TianQin for more values of $P_0/m_0$. 
One can see that for a given detector, the greater the value of $P_0/m_0$, the smaller the value of $m_0$ with which the maximum horizon is reached. 
In addition, the maximum horizon for different values of $P_0/m_0$ does not grow monotonically with $P_0/m_0$.
For example, the maximum horizon for $P_0/m_0=1$ is apparently smaller than those for $P_0/m_0=0.5$ and $P_0/m_0=0.75$.

Secondly, we consider the expected precision of the parameters, $\theta^a\in\{m_0, P, D_L\}$, where $m_0$ is the rest mass of the small black holes, $P$ is the final momentum after the two stages of acceleration, and $D_L$ is the luminosity distance of the source,
\bea \delta\vartheta^a\equiv\frac{\Delta\vartheta^a}{\vartheta^a}\,,\quad 
\Delta\vartheta^a \approx \sqrt{(\Gamma^{-1})^{aa}}\,,\eea 
where $\Gamma^{-1}$ is the inverse of the \ac{FIM},
\bea \Gamma_{ab}=\Big(\frac{\pd h}{\pd\vartheta^a}\Big|\frac{\pd h}{\pd\vartheta^b}\Big)\,.\eea

\begin{figure}[htbp!]
    \centering
    \includegraphics[width=1\linewidth]{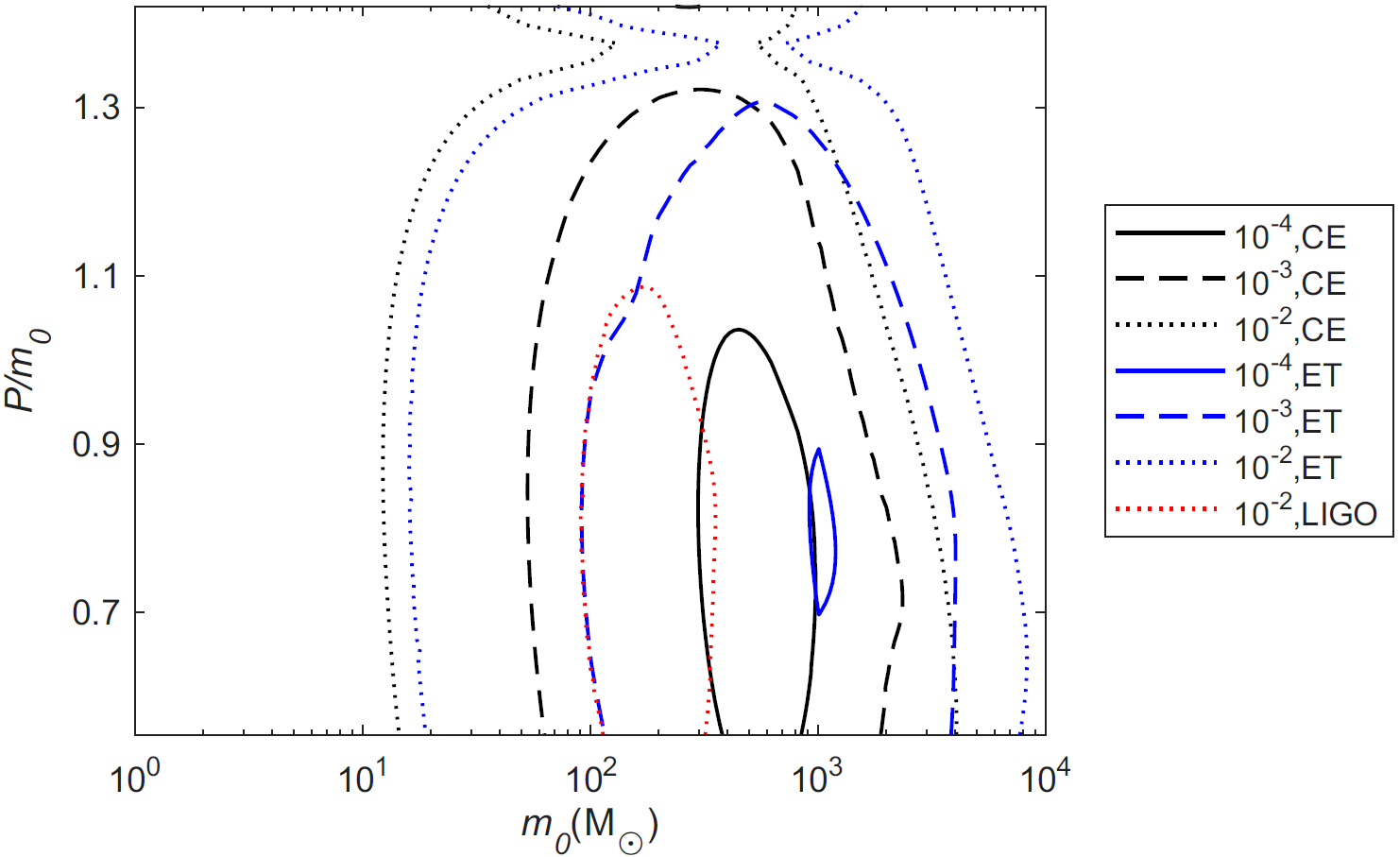}
    \caption{The projected relative precision $\delta m_0$ for ground-based detectors, CE, ET and LIGO. The source is assumed to be at $z = 0.09$.}
    \label{fig:dltm-g}
\end{figure}

\begin{figure}[htbp!]
    \centering
    \includegraphics[width=1\linewidth]{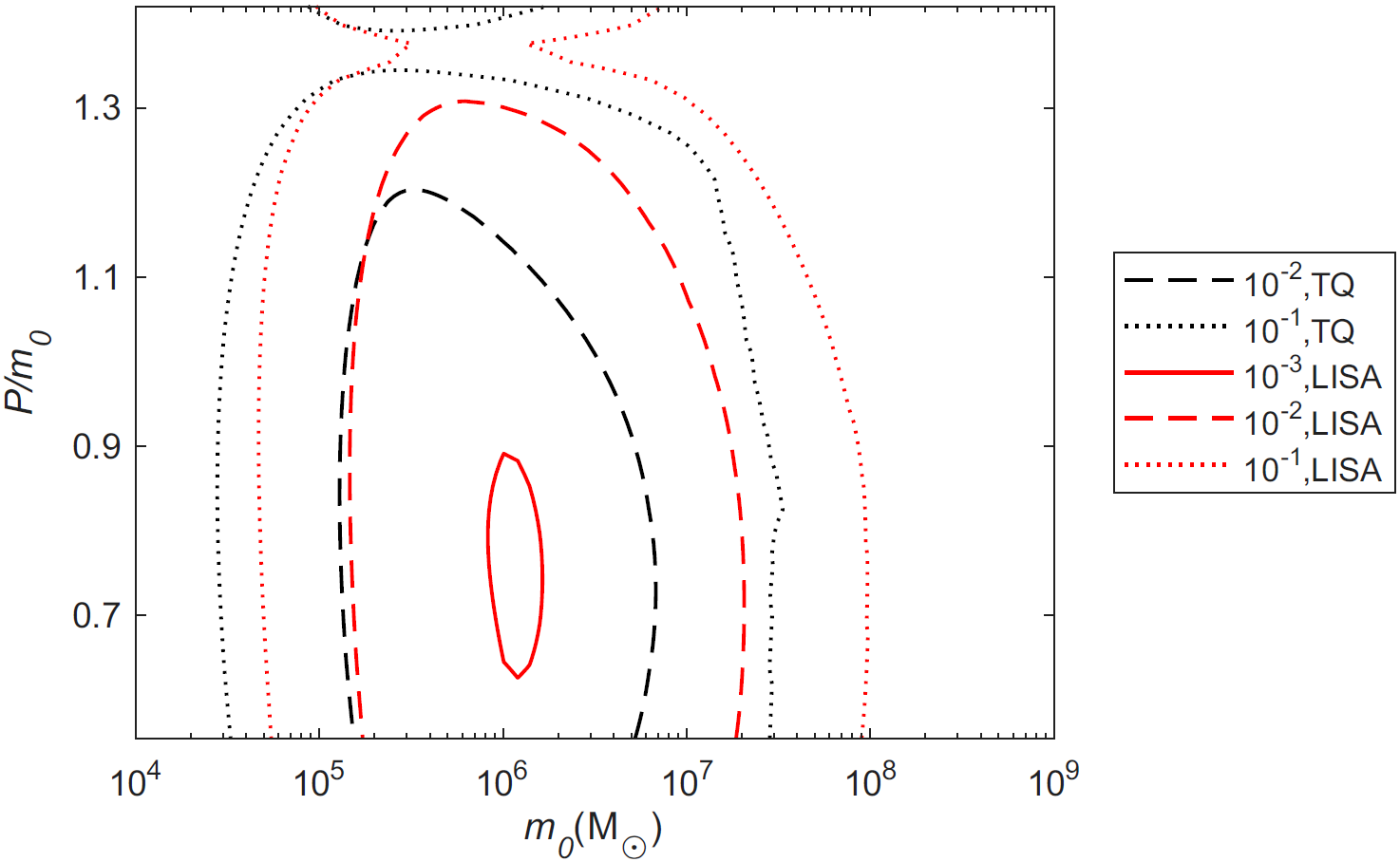}
    \caption{The projected relative precision $\delta m_0$ for space-based detectors, TianQin and LISA. The source is assumed to be at $z = 2$.}
    \label{fig:dltm-s}
\end{figure}

The projected relative precision for measuring the rest mass $m_0$ is plotted in Fig. \ref{fig:dltm-g} and Fig. \ref{fig:dltm-s}.
For \ac{CE} and \ac{ET}, a relative precision of the order $\cO(10^{-4})$ is possible for part of the parameter space, while for LIGO, a relative precision of the order $\cO(10^{-2})$ is possible.
For TianQin and LISA, a relative precision of the order $\cO(10^{-2})$ is possible for a significant part of the parameter space.
For LISA, it is even possible to get a relative precision of the order $\cO(10^{-3})$.
For both ground- and space-based detectors, there is a drastic decrease in the relative precision near $P/m_0 \approx 1.35$.
This might have been caused by the degeneracy between the mass and the momentum near this particular value of the momentum.
However, more research is needed to determine the exact cause.

\begin{figure}[htbp!]
    \centering
    \includegraphics[width=1\linewidth]{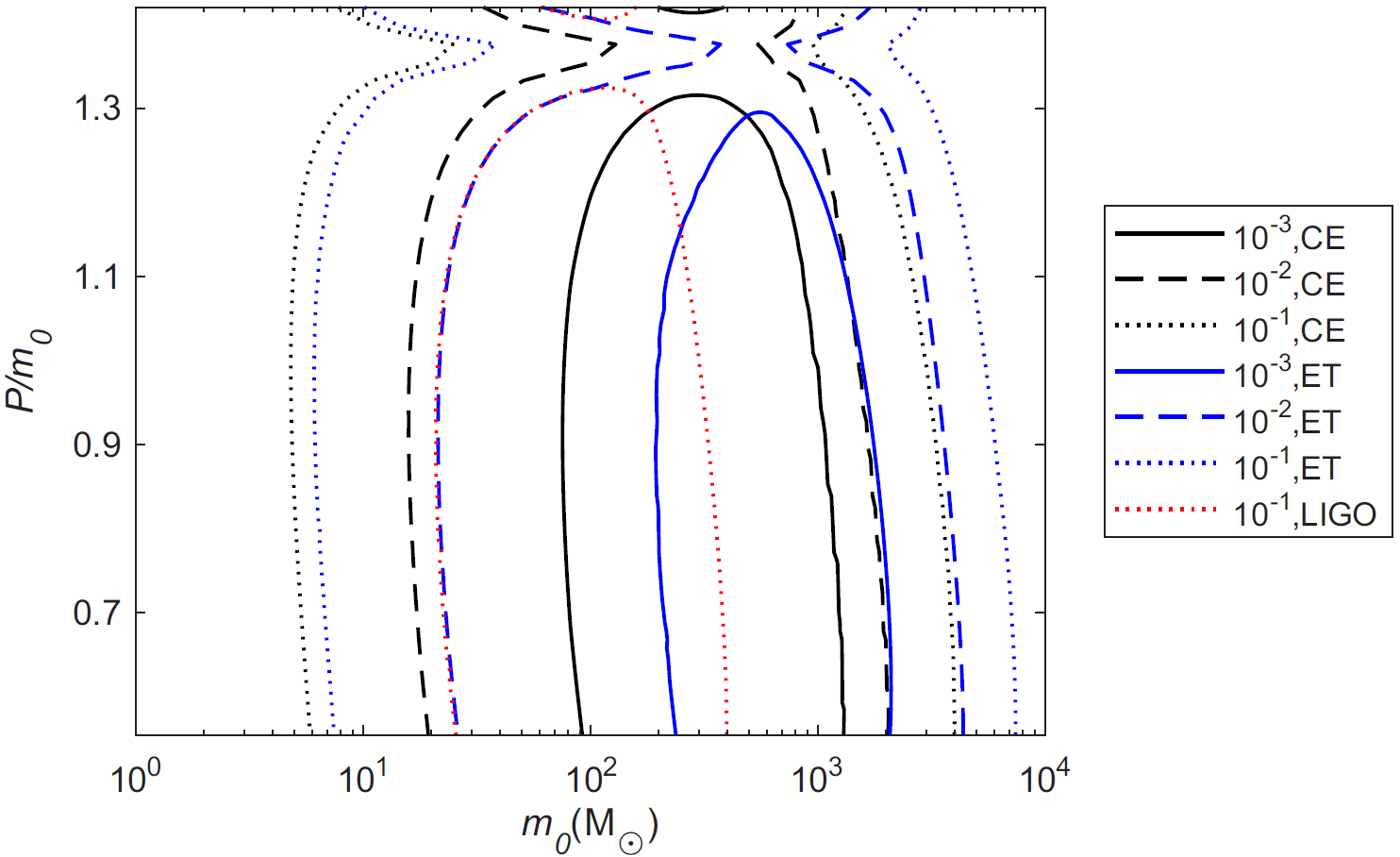}
    \caption{The projected relative precision $\delta P$ for ground-based detectors, CE, ET and LIGO. The source is assumed to be at $z = 0.09$.}
    \label{fig:dltp-g}
\end{figure}

\begin{figure}[htbp!]
    \centering
    \includegraphics[width=1\linewidth]{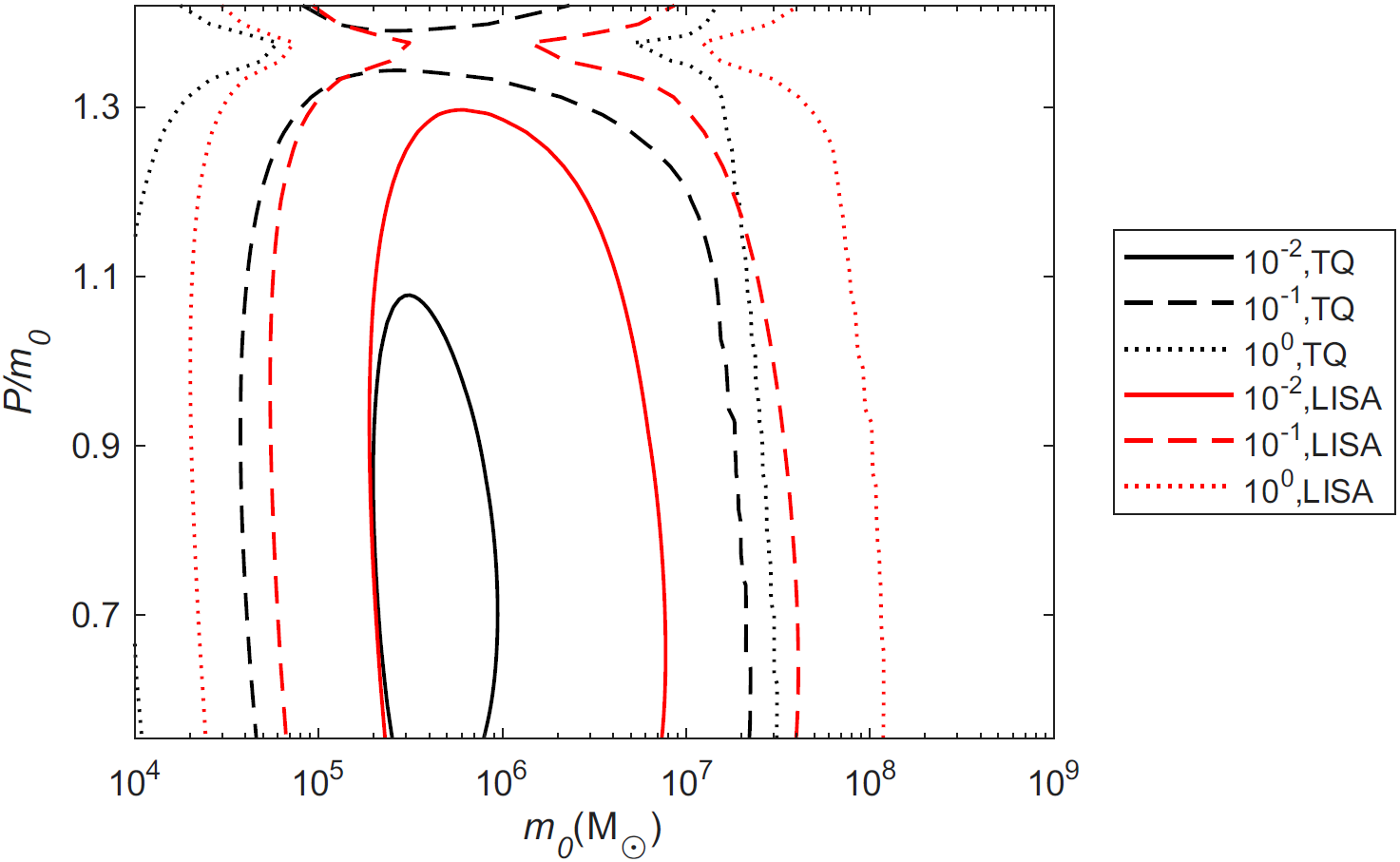}
    \caption{The projected relative precision $\delta P$ for space-based detectors, TianQin and LISA. The source is assumed to be at $z = 2$.}
    \label{fig:dltp-s}
\end{figure}

The projected relative precision for measuring the momentum $P$ is plotted in Fig. \ref{fig:dltp-g} and Fig. \ref{fig:dltp-s}.
For \ac{CE} and \ac{ET}, a relative precision of the order $\cO(10^{-3})$ is possible for part of the parameter space, while for LIGO, a relative precision of the order $\cO(10^{-1})$ is possible.
For TianQin and LISA, a relative precision of the order $\cO(10^{-2})$ is possible for part of the parameter space.
Similarly to the case of $\delta m_0$, for both ground- and space-based detectors, there is a drastic decrease in the relative precision $\delta P$ near $P/m_0 \approx 1.35$.
Again, this might have been caused by the degeneracy between the mass and the momentum near this particular value of the momentum, but more research is needed to determine the exact cause.

\begin{figure}[htbp!]
    \centering
    \includegraphics[width=1\linewidth]{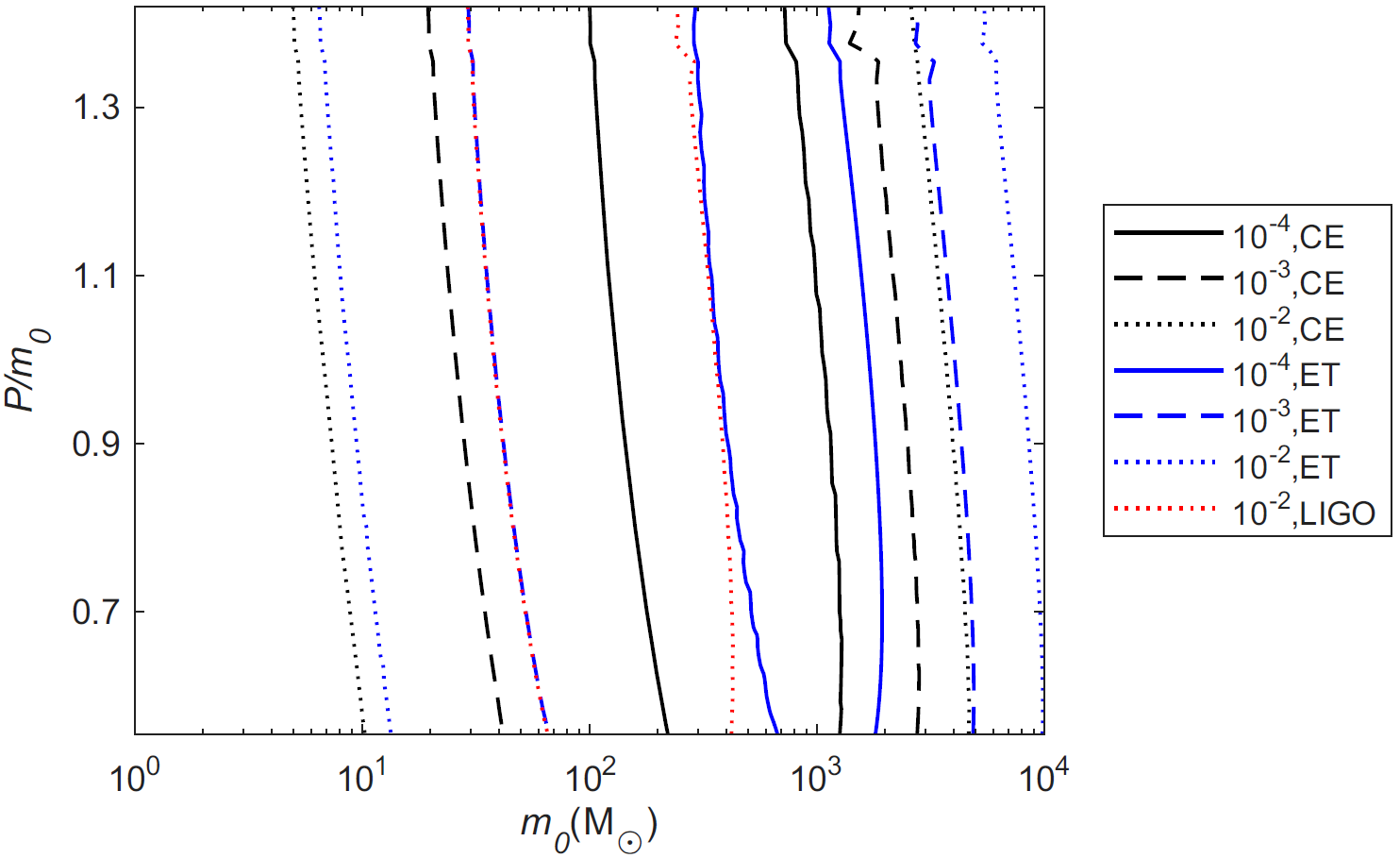}
    \caption{The projected relative precision $\delta D_L$ for ground-based detectors, CE, ET and LIGO. The source is assumed to be at $z = 0.09$.}
    \label{fig:dltDL-g}
\end{figure}

\begin{figure}[htbp!]
    \centering
    \includegraphics[width=1\linewidth]{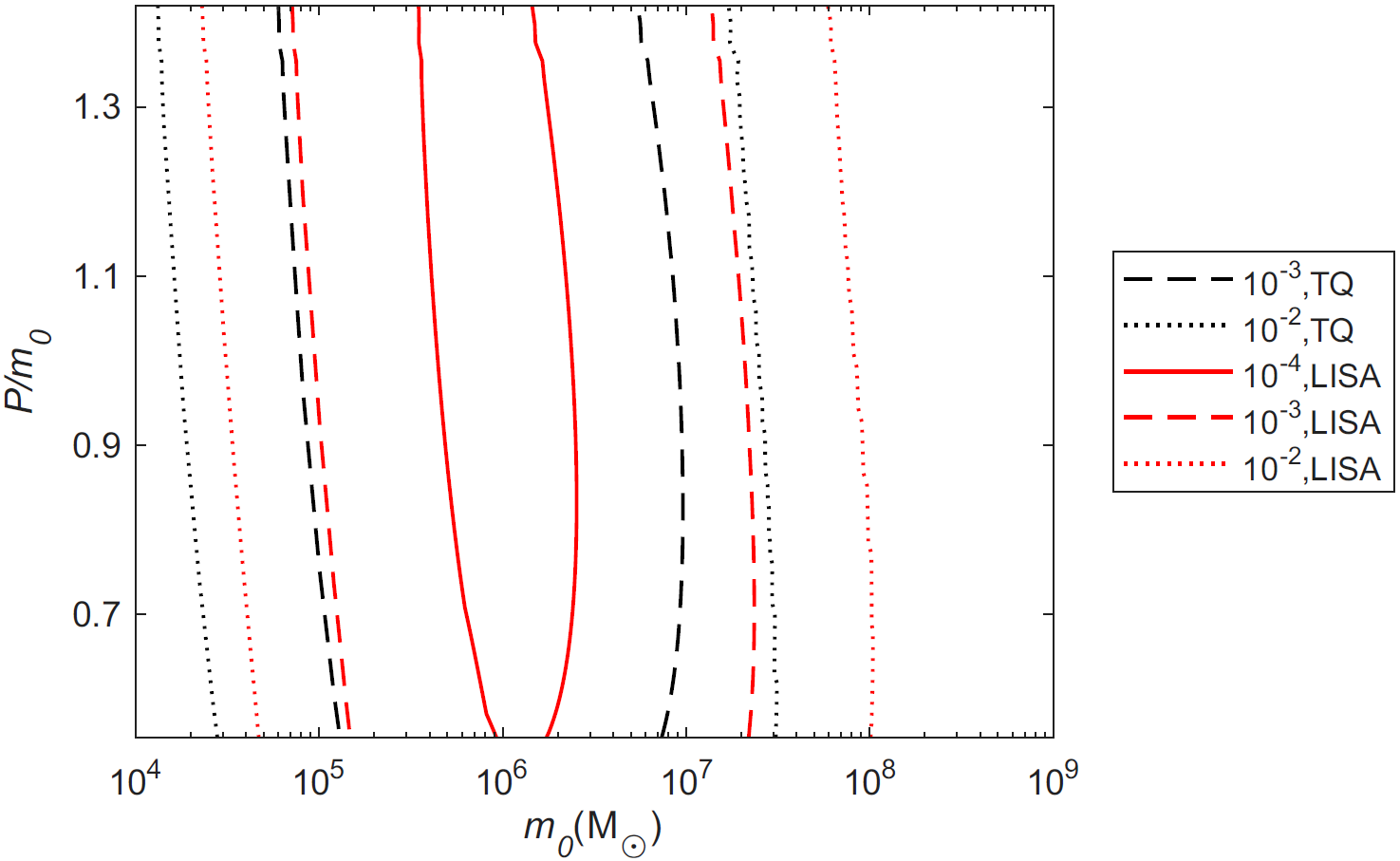}
    \caption{The projected relative precision $\delta D_L$ for space-based detectors, TianQin and LISA. The source is assumed to be at $z = 2$.}
    \label{fig:dltDL-s}
\end{figure}

The projected relative precision for measuring the momentum $D_L$ is plotted in Fig. \ref{fig:dltDL-g} and Fig. \ref{fig:dltDL-s}.
For \ac{CE} and \ac{ET}, a relative precision of the order $\cO(10^{-4})$ is possible for part of the parameter space, while for LIGO, a relative precision of the order $\cO(10^{-2})$ is possible.
For TianQin and LISA, a relative precision of the order $\cO(10^{-3})$ is possible for part of the parameter space.
For LISA, it is even possible to get a relative precision of the order $\cO(10^{-4})$.
In contrast to the cases of $\delta m_0$ and $\delta P$, there is not much decrease in the relative precision $\delta D_L$ near $P/m_0 \approx 1.35$.

\section{Summary}
\label{sec:sum}

In this paper, we have studied the intriguing scenario of \acp{RBHC} taking place in a \ac{MBH} background.
The two small black holes are expected to undergo two stages of acceleration before collision: the first caused by the background \ac{MBH} and the second caused by their own mutual attraction.

Focusing on the special cases that the two small black holes move in the equitorial plane of the \ac{MBH} and that the collision occurs at the periastron of one of the small black holes, we find that the largest boost can be achieved when the background \ac{MBH} is not rotating.
When the two small black holes are on a common circular orbit but move in opposite directions, the largest boost is about $P/m_0\approx0.94$ when the two small black holes are on the \ac{ISCO} of the \ac{MBH}.
When the two small black holes are on a common eccentric orbit but move in opposite directions, the largest boost is about $P/m_0\approx1.42$ when the eccentricity of the orbit approaches 1.
One can get the same largest boost $P/m_0\approx1.42$ if the two small black holes are on the same lowest scattering trajectory (assuming $E=1$) of the \ac{MBH} but move in opposite directions.

We use \ac{CLAP} to calculate the waveform of the collision.
We find that the difference caused by different initial boosts is most noticeable near the peaks and troughs of the wavefroms.
In addition, we find that for $P/m_0\lesssim1$, the radiated energy in the (4,0) and (6,0) modes is several orders of magnitude smaller than that in the (2,0) mode.
It will be important to find out if more realistic collision scenarios, such as unequal masses, non-zero spins, and a non-zero impact parameter for the small black holes, can help significantly increase the relative contribution of the higher modes.

We have studied the capability of several current and future detectors in detecting the \ac{RBHC} signals, including TianQin, LISA (Taiji), \ac{CE}, \ac{ET} and LIGO.
For future space-based detectors, the detection horizon can reach redshift $z\approx8$ (TianQin) and $z\approx14$ (LISA).
For future ground-based detectors, the detection horizon can reach the redshift $z\approx5$. 
For LIGO (assuming O4 sensitivity),  the detection horizon can reach the redshift $z\approx0.5$.
For TianQin, the expected relative precision of parameter estimation can reach the order $\cO(10^{-2})\,$, $\cO(10^{-2})\,$, $\cO(10^{-3})\,$ for $(m_0, P, D_L)$, respectively.
With favorable source parameters, LISA can achieve the relative precision of parameter estimation at the order $\cO(10^{-3})\,$, $\cO(10^{-2})\,$, $\cO(10^{-4})\,$ for $(m_0, P, D_L)$, respectively.
For future ground-based detectors, the expected relative precision of parameter estimation can reach the order $\cO(10^{-4})\,$, $\cO(10^{-3})\,$, $\cO(10^{-4})\,$ for $(m_0, P, D_L)$, respectively.
LIGO is about 2 orders of magnitude worse than \ac{CE} or \ac{ET}.

\section*{Acknowledgments}

The authors thank Ulrich Sperhake for a helpful communication.
The work has been supported in part by the National Key Research and Development Program of China (Grant No. 2023YFC2206700), the Natural Science Foundation of China (Grants No. 12261131504), and the Fundamental Research Funds for the Central Universities, Sun Yat-sen University.

\bibliographystyle{apsrev4-1}
\bibliography{ref}
\end{document}